\documentstyle[preprint,aps,eqsecnum,epsfig]{revtex}
\begin{document}
\draft
\preprint{\vbox{\baselineskip16pt
\hbox{KEK-TH-474}
\hbox{SNUTP 96-019}
\hbox{YUMS 96-009}
}}
\title{Color-Octet Contributions in\\
the Associate $J/\psi + \gamma$ Hadroproduction }
\author{C. S.~~Kim\footnote{kim@cskim.yonsei.ac.kr,~~cskim@kekvax.kek.jp}
}
\address{
Department of Physics, Yonsei University, Seoul 120-749, Korea}
\address{
Theory Division, KEK, Tsukuba, Ibaraki 305, Japan}
\author{Jungil~~Lee\footnote{jungil@phya.snu.ac.kr}
and H. S.~~Song\footnote{hssong@physs.snu.ac.kr}}
\address{
Department of Physics and Center for Theoretical Physics
\\ Seoul National University,
Seoul 151-742, Korea  }
\maketitle
\vspace*{-1.0cm}

\begin{abstract}
Color-octet contributions to the associate $J/\psi+\gamma$ 
hadroproduction are studied  in detail, and found to be negligible,
compared to the ordinary color-singlet contribution.
Within the color-singlet model the $J/\psi+\gamma$ production in the leading 
order is possible only through gluon-gluon fusion process.
Therefore, the associate $J/\psi+\gamma$ hadroproduction remains 
to be  useful as a clean channel to probe the gluon distribution
inside proton, to study  heavy quarkonia production mechanism,
and to find proton's spin structure.
\end{abstract}


\section{Introduction}

Up to very recently the conventional color-singlet model \cite{berjon,baier} 
had been used as only possible mechanism to
describe the production and decay of the heavy quarkonium 
such as $J/\psi$ and $\Upsilon$. An artificial $K$-factor ($\sim$ 2--5) 
was first introduced to  compensate the gap between the theoretical 
prediction and experimental data.
The uncertainties related to this large $K$-factor are the following:
We don't know precisely the correct  normalization of the bound state 
wave function, the possible next-to-leading order contributions, 
the mass of the heavy quark inside the bound state, {\it etc}.
However, even with this large $K$-factor some experimental data 
are found to be difficult to describe. For the direct $J/\psi$ production 
with large $p_{_T}$ in $p\overline p$ collisions,
the dominant mechanism has been found to be through 
final parton fragmentation \cite{frag_bryu}.
As applications of this fragmentation mechanism,
various studies of prompt charmonium production at the Tevatron collider
\cite{frag_brdon,frag_rs,frag_cacci} have been carried out,
and the CDF data on the prompt $J/\psi$ production \cite{CDF94} 
qualitatively meet these theoretical predictions.
Nevertheless, the $\psi^\prime$ production rate at CDF 
is about 30 times larger than the theoretical predictions,
which is the so-called $\psi^\prime$ anomaly,
even after considering the fragmentation mechanisms of
$g\rightarrow \psi^\prime$ and $c\rightarrow \psi^\prime$. 
As a scenario to resolve this $\psi^\prime$ anomaly, 
the color-octet production mechanism was proposed \cite{psi_prime}.
By using this idea, heavy quarkonium (charmonium) hadroproduction
through the color-octet $(c \overline{c})_8$ pair in various partial
wave  states $(^{2S+1}L_J)$ has been studied 
in addition to the color-octet gluon fragmentation approach
\cite{pchoone,pchotwo}.
We note that the inclusive $\Upsilon$ production at the
Tevatron also shows an excess of the data over theoretical estimates 
based on perturbative QCD and the color-singlet model 
\cite{vaia}.
In this case, the $p_{_T}$ of the $\Upsilon$ is not so high, so that the gluon
fragmentation picture may not be a good approximation any more.
 
Since it has been proposed that the color-octet mechanism 
can resolve the $\psi^{'}$ anomaly at the
Tevatron collider, it is quite important to test this mechanism at  other 
high energy heavy quarkonium production processes. 
Up to now, the following 
processes have been theoretically considered: inclusive 
$J/\psi$ production at the Tevatron collider
and at fixed target experiments \cite{pchoone,pchotwo,fleming1},
spin alignment of leptons to the decayed $J/\psi$ \cite{wise}, 
the polar angle distribution of the $J/\psi$ in
$e^+ e^-\to J/\psi +X$ \cite{cleo_br},
inclusive $J/\psi$ productions in $B$ meson decays \cite{jungilb,jungilep}  
and in $Z^0$ decays at LEP \cite{cheungz,pchoz,jungilz},
$J/\psi$ photoproduction \cite{jungilep,kramer2,fleming2,kramer_new},
and color-octet $J/\psi$ production in
$\Upsilon$ decays \cite{upsilon}. 
For more details on recent progress, see Ref. \cite{br_review},
which summarizes the theoretical developments on the quarkonium production.
Recently, 
the helicity decomposition method in NRQCD factorization formalism
was developed by Braaten and Chen~\cite{bra_chen}.
With this method, polarized $J/\psi$ production in $B$ decay was
considered in Ref.~\cite{pol_b}.
In Ref.~\cite{bra_chen,bene_roth},
it was pointed out that there are  interference terms of different 
$^3P_J$ contributions in polarized $J/\psi$ production.

In this work, we investigate the color-octet mechanism
for associate $J/\psi+\gamma$ production in hadronic collisions.
The associate $J/\psi+\gamma$ production  has been first
proposed as a clean channel to probe the gluon distribution
inside the proton or photon \cite{psigamma_kimdrees}, and then
to study the heavy quarkonia production mechanism \cite{psigamma_kimreya}
as well as to investigate the proton's spin structure \cite{psigamma_pol}.
If the  color-octet  mechanism gives a significant contribution 
to this process, the merit of this process to probe the 
gluon distribution {\it etc.}
would be decreased or one should find suitable cuts to get rid of this
new contribution.
Associate $J/\psi+\gamma$ production has also been studied
as a significant QCD background to the decay of heavier $P$-wave 
charmonia ($\chi_{cJ}(P)$) in fixed target experiments \cite{e705,isr}
within the color-singlet model \cite{psigamma_mass}.
In Ref. \cite{psigamma_frag}, the fragmentation contribution from
$p + \bar p \rightarrow (c~{\rm or}~g) + \gamma + X \rightarrow J/\psi
+ \gamma + X$
at Tevatron energies ($\sqrt{s}$=1.8~TeV) is studied, and 
these fragmentation channels are found to be significantly 
suppressed compared to the conventional
color-singlet gluon-gluon fusion process.

In Section 2, we explain in detail the method to deal with
inclusive heavy quarkonium production, 
following the nonrelativistic-QCD (NRQCD) factorization formalism given 
by Bodwin, Braaten and Lepage (BBL) \cite{BBL}. 
A summary of the kinematics related to the $p_{_T}$ and $M_{J/\psi+\gamma}$ 
distributions is also given in this Section.
In Section 3, we discuss in detail the $p_{_T}$ and 
$M_{J/\psi+\gamma}$ distributions in the hadronic $J/\psi+\gamma$ 
production, and compare the color-octet contribution 
with the color-singlet gluon-gluon fusion contribution.
Section 3 also contains our conclusions.

\section{Hadroproduction of $J/\psi + \gamma$}
\subsection{General Discussions}

Now we consider the associate production of 
a $J/\psi$ and a photon at hadronic colliders and fixed target 
experiments, including the new contribution from the color-octet mechanism
in addition to the usual color-singlet contribution. 
The only possible subprocess in the framework of the color-singlet model 
in the leading order is through gluon-gluon fusion,
\begin{eqnarray}
g+g\rightarrow \gamma+(c\overline{c})(^{3}S^{(1)}_{1})(\rightarrow J/\psi).
~~~~~({\rm See ~Fig.~\ref{fig:fusion}})
\label{eq:gluonfusion}
\end{eqnarray}
This gluon-gluon fusion process for the associate $J/\psi+\gamma$ production 
is again possible within the color-octet mechanism.
(See Fig. \ref{fig:fusion}).
However,  in the leading order color-octet contributions
there also exist many more subprocesses, which are 
\begin{eqnarray}
g+g&\rightarrow& \gamma+(c\overline{c})(^{1}S^{(8)}_{0})(\rightarrow J/\psi),
\label{eq:gluon_1s0}\\
g+g&\rightarrow& \gamma+(c\overline{c})(^{3}P^{(8)}_{J})(\rightarrow J/\psi),
\label{eq:gluon_3pj}\\
q+\overline{q}
&\rightarrow& \gamma+(c\overline{c})(^{1}S^{(8)}_{0})(\rightarrow J/\psi),
\label{eq:quark_1s0}\\
q+\overline{q}&\rightarrow& 
\gamma+(c\overline{c})(^{3}P^{(8)}_{J})(\rightarrow J/\psi),
\label{eq:quark_3pj}\\
q+\overline{q}&\rightarrow& 
\gamma+(c\overline{c})(^{3}S^{(8)}_{1})(\rightarrow J/\psi).
\label{eq:quark_3s1}
\end{eqnarray}
These subprocesses are possible through the effective vertices of
\begin{equation}
\gamma+g\rightarrow (c\overline{c})(
^1S_0^{(8)}
~~{\rm or}~~
^3P_J^{(8)}),
\label{eq:agp}
\end{equation}
which is shown in Fig. \ref{fig:agp}, and
\begin{equation}
q+\overline{q}\rightarrow (c\overline{c})(
^3S_1^{(8)}),
\label{eq:qq3s1}
\end{equation}
which is shown in Fig. \ref{fig:qqp}.
We note that the $(^3S_1^{(8)})$ channel in Eq. (\ref{eq:agp}) 
is vanishing, even though the gluon is not on its mass shell. 
The Feynman diagram for the gluon initiated subprocesses,
Eqs. (\ref{eq:gluon_1s0}, \ref{eq:gluon_3pj}),
is shown in Fig. \ref{fig:octet22} (a).
For the case of quark initiated subprocesses, there are two kinds of 
diagrams;  The Feynman diagram for quark initiated subprocesses
Eqs. (\ref{eq:quark_1s0}, \ref{eq:quark_3pj}),
due to the effective vertex Eq. (\ref{eq:agp}), 
is shown in Fig. \ref{fig:octet22} (b).
The other diagrams, which are possible through 
the effective vertex, Eq. (\ref{eq:qq3s1}),  
are shown in Fig. \ref{fig:qq3s1}.

We first perform a naive power counting analysis of those various
subprocesses.
Free particle amplitude-squared for the
color-singlet gluon-gluon fusion process (\ref{eq:gluonfusion})
is $[{\cal O}(\alpha \alpha_{s}^2)]$.
In the NRQCD framework \cite{BBL}, 
the color-octet ($Q\overline Q$) states can also form a physical 
$J/\psi$ state with dynamical gluons inside the quarkonium 
with wavelengths much larger than the characteristic size of the
bound state ($\sim$$1/(M_Q v_Q)$).
In Coulomb gauge, which is a natural gauge for analyzing heavy quarkonium,
these dynamical gluons enter into the Fock state decomposition
of physical state of $J/\psi$ as
\begin{eqnarray}
|J/\psi\rangle=&&
 {\cal O}(1    )|(Q\overline{Q})(^3S_1^{(1)})\rangle
+{\cal O}(v_Q  )|(Q\overline{Q})(^3P_J^{(8)})g\rangle\nonumber\\
&+&
 {\cal O}(v_Q^2)|(Q\overline{Q})(^3S_1^{(1,8)})gg\rangle
+{\cal O}(v_Q^2)|(Q\overline{Q})(^1S_0^{(8)})g\rangle+...~.
\end{eqnarray}
In general, a state
$(Q\overline{Q})(^{2S+1}L_J)$ can make a transition to
$(Q\overline{Q})(^{2S+1}(L\pm 1)_J)$, or more specifically
$(Q\overline{Q})(^3P_J^{(8)})\rightarrow (Q\overline{Q})(^3S_1^{(1)})$
through the emission of a soft gluon (chromo-electric dipole transition),
and it is an order of $v_Q$ suppressed compared to 
the color-singlet hadronization.
For the case of
chromo-magnetic dipole transition, such as 
$(Q\overline{Q})(^1S_0^{(8)})\rightarrow (Q\overline{Q})(^3S_1^{(1)})$,
it is suppressed by $v_Q$ at the amplitude level.
If we also consider the fact that
the $P$ wave state is $M_Qv_Q$ order higher than 
$S$ wave state in the amplitude level,
one can naively estimate that the transitions
\begin{eqnarray}
(Q\overline{Q})(^3P_J^{(8)})\rightarrow J/\psi+X~~~{\rm and}~~~
(Q\overline{Q})(^1S_0^{(8)})\rightarrow J/\psi+X
\end{eqnarray}
are commonly $v_Q^4$ order  suppressed at the amplitude-squared level, 
compared to the transition
\begin{eqnarray}
(Q\overline{Q})(^3S_1^{(1)})\rightarrow J/\psi~.
\end{eqnarray}
Since these color-octet processes have the same order  
${\cal O}(\alpha \alpha_s^2)$
as that of the color-singlet gluon-gluon fusion process in free particle 
scattering amplitude,
the color-octet subprocesses are order $v_Q^4$ suppressed 
compared to the color-singlet gluon-gluon fusion subprocess.
 
However, such analyses are only naive power counting,
and  do not guarantee that the 
color-octet contributions are suppressed compared to the color-singlet
gluon-gluon fusion contribution all over the allowed kinematical region.
If we consider inclusive $J/\psi$ photoproduction via 
$2\rightarrow 2$ subprocesses, 
as shown in Refs. \cite{jungilep,kramer2},
color-octet contributions dominate in some kinematical region,
even if the naive power counting predicts the suppression 
of color-octet contributions compared to the color-singlet one.
In this respect, it is worthwhile to investigate in detail how much the
color-octet mechanism contributes to the $J/\psi+\gamma$ 
hadroproduction.

In order to predict numerically the physical production rate,
we need to know a few nonperturbative parameters characterizing the
fragmentation of the color-octet objects into the physical 
color-singlet $J/\psi$. 
These nonperturbative matrix elements for the color-octet operators 
have not been determined completely yet.
After fitting the inclusive $J/\psi$ production at the Tevatron collider, 
using the usual color-singlet $J/\psi$ production, 
the cascades production from $\chi_{c}(1P)$, and the new
color-octet contributions,  the authors of Ref.~\cite{pchotwo} have
determined
\begin{eqnarray}
\langle 0 | O_{8}^{\psi} (^{3}S_{1}) | 0 \rangle   &=&   (6.6 \pm 2.1)
\times 10^{-3}~{\rm GeV}^3,
\\
\frac{\langle 0|{\cal O}_{8}^{\psi}({^3P_{0}})|0\rangle}{M_c^2}
     +\frac{\langle 0|{\cal O}_{8}^{\psi}({^1S_{0}})|0\rangle}{3}
&=&(2.2\pm 0.5)\times 10^{-2}~{\rm GeV}^3,
\label{eq:fit}
\end{eqnarray}
with $M_{c} = 1.48$ GeV.
Although the numerical values of two matrix elements, $\langle 0|
{\cal O}_{8}^{\psi}(^3P_0)|0\rangle$ and $\langle 0|{\cal O}_{8}^{\psi}
(^1S_0)|0\rangle$, are not separately known 
in Eq.~(\ref{eq:fit}), one can still
extract some useful information from them. Assuming both of the color-octet
matrix elements in Eq.~(\ref{eq:fit}) 
are positive definite, then one has\cite{jungilep}
\footnote{The matrix element $\langle 0|{\cal O}_{8}^{\psi}({^3P_0})|0\rangle$
can be negative whereas  $\langle 0|{\cal O}_{8}^{\psi}({^1S_0})|0\rangle$
is always positive definite~\cite{pol_b}.
Here we choose these ranges just for simplicity.
}
\begin{eqnarray}
0 < \langle 0|{\cal O}_{8}^{\psi}({^1S_0})|0\rangle < (6.6 \pm 1.5)
\times 10^{-2}~{\rm GeV}^3,
\\
0 < { \langle 0|{\cal O}_{8}^{\psi}({^3P_0})|0\rangle \over M_{c}^2}
<  (2.2 \pm 0.5) \times 10^{-2}~{\rm GeV}^3.
\label{eq:max}
\end{eqnarray}
These inequalities could provide us with a few predictions on various
quantities related to inclusive $J/\psi$ productions in other
processes, and also enable us to test the idea of color-octet
mechanism in the associate $J/\psi+\gamma$ production process.

\subsection{Kinematics}

Let us consider the process
\begin{equation}
a(p_1)_{/p(P_1)} + b(p_2)_{/\overline{p}(P_2)} 
\rightarrow J/\psi(P)+\gamma(k).
\end{equation}
in $p \overline p$ (or alternatively $p p$) collisions.
We can express the momenta of the incident hadrons $(p,~\bar p)$ and 
partons $(a,~b)$ in the $p \overline{p}$ CM frame as
\begin{eqnarray}
P_1=\frac{\sqrt{s}}{2}(1,+1,0), ~~&&~~ p_1=\frac{\sqrt{s}}{2}(x_1,+x_1,0),\\
P_2=\frac{\sqrt{s}}{2}(1,-1,0), ~~&&~~ p_2=\frac{\sqrt{s}}{2}(x_2,-x_2,0),
\end{eqnarray}
where the first component is the energy, the second is longitudinal momentum,
and the third is the transverse component of the particle's momentum.
The variables $x_1$ and $x_2$ are the momentum fractions of the partons.
The momenta of the outgoing particles are given by,
\begin{eqnarray}
 P=(E^\psi,P_L^\psi,+P_T)&=&( M_T\cosh y^\psi,M_T\sinh y^\psi,+P_T),\\
 k~=(E^\gamma,P_L^\gamma,-P_T)&=&(~P_T\cosh y^\gamma,~P_T\sinh y^\gamma,-P_T),
\end{eqnarray}
where $P_T$ is the common transverse momentum of the outgoing particles,
$M_T$ is the transverse mass of the outgoing $J/\psi$,
and $y^\psi$ (or $y^\gamma$) is the rapidity of $J/\psi$ (or $\gamma$).
Some useful relations among the above variables are:
\begin{eqnarray}
E^\psi+P_L^\psi=M_Te^{+y^\psi},
~~&&~~E^\psi-P_L^\psi=M_Te^{-y^\psi},
\\
E^\gamma+P_L^\gamma=~P_Te^{+y^\gamma},
~~&&~~E^\gamma-P_L^\gamma=~P_Te^{-y^\gamma}.
\end{eqnarray}
The Mandelstam variables are defined respectively as
\begin{eqnarray}
s&=&(P_1+P_2)^2,\\
\hat{s}=(p_1+p_2)^2&=& 2x_1x_2P_1\cdot P_2=x_1x_2s,\\
\hat{t}=(p_1-~k )^2&=&-x_1\sqrt{s}P_Te^{-y^\gamma}
        = (p_2-P  )^2=M_\psi^2-x_2\sqrt{s}M_Te^{+y^\psi},\\
\hat{u}=(p_2-~k )^2&=&-x_2\sqrt{s}P_Te^{+y^\gamma}
        = (p_1-P  )^2=M_\psi^2-x_1\sqrt{s}M_Te^{-y^\psi},\\
M_\psi^2&=&\hat{s}+\hat{t}+\hat{u}.
\end{eqnarray}
The last equation, representing the energy-momentum conservation,
leads to  
\begin{eqnarray}
\sqrt{s}M_T(x_2e^{y^\psi}  +x_1e^{-y^\psi})&=&\hat{s}+M_\psi^2,\nonumber\\
\sqrt{s}P_T~(x_2e^{y^\gamma}+x_1e^{-y^\gamma})&=&\hat{s}-M_\psi^2.
\end{eqnarray}
After introducing dimensionless variables,
\begin{equation}
x_T=2P_T/\sqrt{s}
,\hskip.7cm
\overline{x}_T=2M_T/\sqrt{s}
\hskip.5cm
{\rm and}
\hskip.5cm
\tau=M^2_\psi/s,
\end{equation}
we can solve the energy momentum relation
to get $x_2$ in terms of other variables as
\begin{equation}
x_2=\frac{x_1\overline{x}_Te^{-y^\psi}  -2\tau}
         {2x_1-\overline{x}_Te^{+y^\psi}}
\hskip .5cm
{\rm or}
\hskip .5cm
x_2=\frac{x_1      x_Te^{-y^\gamma}+2\tau}
         {2x_1-x_Te^{+y^\gamma}}.
\end{equation}
We also obtain the relations for rapidities
of the outgoing particles in terms of $x_1$, $x_2$, $x_T$, 
$\overline{x}_T$ and $\tau$ as following
\begin{eqnarray}
{\rm exp}({y^\psi}) &=&  
\frac{(x_1x_2 + \tau) + \sqrt{(x_1x_2 + \tau)^2 -x_1x_2 \overline x_T^2}} 
     {x_2 \overline x_T},\nonumber\\
{\rm exp}({y^\gamma}) &=&  
\frac{(x_1x_2 - \tau) - \sqrt{(x_1x_2 - \tau)^2 -x_1x_2      x_T^2}} 
     {x_2      x_T}.
\end{eqnarray}
In order to get the distributions in the
invariant mass $M_{J/\psi+\gamma}(\equiv \sqrt{\hat{s}})$ 
and transverse momentum $P_{_T}$ for the process
$g+g \rightarrow J/\psi+\gamma$ process,  
we express the differential cross section as
\begin{eqnarray}
d\sigma
&=& f_{g/p}(x_1,Q^2) f_{g/\overline{p}}(x_2,Q^2)
\frac{d\hat{\sigma}}{d\hat{t}}dx_1dx_2d\hat{t}
\nonumber\\
&=& f_{g/p}(x_1,Q^2) f_{g/\overline{p}}(x_2,Q^2)
\frac{d\hat{\sigma}}{d\hat{t}}
J\left(\frac{x_1x_2\hat{t}}{x_1x_{T}M_{J/\psi+\gamma}}\right)
dx_1dx_TdM_{J/\psi+\gamma}
\nonumber\\
&=& f_{g/p}(x_1,Q^2) f_{g/\overline{p}}(x_2,Q^2)
\frac{d\hat{\sigma}}{d\hat{t}}
J\left(\frac{x_1x_2\hat{t}}{x_1y^\psi P_{T}}\right)
dx_1dy^\psi dP_{T},
\end{eqnarray}
where the corresponding {\it Jacobians} are given by
\begin{eqnarray}
J\left(\frac{x_1x_2\hat{t}}{x_1x_{T}M_{J/\psi+\gamma}}\right)
= \frac{2x_2x_{T}M_{J/\psi+\gamma}}
         {\overline{x}_{T}(x_2e^{+y_\psi}-x_1e^{-y_\psi})}
~~~{\rm and}~~~
J\left(\frac{x_1x_2\hat{t}}{x_1y^\psi P_{T}}\right)
= \frac{4x_1x_2P_T}{2x_1-\overline{x}_{T}e^{+y^\psi}}
\end{eqnarray}
Then the distributions are expressed as
\begin{eqnarray}
\frac{d\sigma}{dM_{J/\psi+\gamma}}
&=&
\int dx_1~dx_{T}
J\left(\frac{x_1x_2\hat{t}}{x_1x_{T}M_{J/\psi+\gamma}}\right)
\frac{d^3\hat{\sigma}}{dx_1dx_2d\hat{t}}~,\\
\frac{d\sigma}{dP_T}
&=&
\int dx_1~dy^\psi 
J\left(\frac{x_1x_2\hat{t}}{x_1y^\psi P_T}\right)
\frac{d^3\hat{\sigma}}{dx_1dx_2d\hat{t}}~.
\end{eqnarray}
And the allowed regions of the variables are given by
\begin{eqnarray}
\frac{M_{J/\psi+\gamma}^2}{s} \le x_1 \le 1,\nonumber \\
\hat{s}=M_{J/\psi}^2=x_1x_2s \ge M_{J/\psi}^2,\nonumber\\
0 \le x_T \le \frac{(x_1x_2-\tau)}{\sqrt{x_1x_2}}.
\end{eqnarray}

\subsection{The Nonrelativistic-QCD (NRQCD) Factorization Formalism}

First we consider the general method to get the NRQCD cross section
for the process
$a+b\rightarrow (Q\overline{Q})(^{2S+1}L^{(1,8)}_J)(\rightarrow  H)+c$,
where $H$ is the final state heavy quarkonium and 
$(Q\overline{Q})(^{2S+1}L^{(1,8)}_J)$
is the intermediate  $(Q\overline{Q})$ pair which has the corresponding
spectroscopic state.
{}From now on, we use the subscript $n$ 
to represent the spectroscopic $(Q\overline Q)$ state of 
$(^{2S+1}L_J^{(n=1,8)})$, for simplicity.
Once the on-shell scattering amplitude of the process
${\cal A}(a+b\rightarrow Q+\overline{Q}+c)$ is given,
we can expand the amplitude in terms of relative momentum $q$ of
the quarks inside the bound state
because the quarks, which make up the bound state, are heavy.
For more details of the method to deal with the heavy quarkonium production 
following the BBL formalism \cite{BBL}, 
we refer to Refs. \cite{pchoone,pchotwo,jungilep}.
The heavy quarkonium $H$ production cross section
$a(p_1)+b(p_2)\rightarrow (Q\overline{Q})_n(P)+c(p_3) \rightarrow  H+c+X $
is given by
\begin{eqnarray}
\frac{d\hat{\sigma}}{d\hat{t}}&&\left(
a(p_1) + b(p_2)\rightarrow (Q\overline{Q})_n(P)+c(p_3)
             \rightarrow  H+c+X
\right)\nonumber\\
&&=
\frac{1}{C_nM_Q}\times
\frac{d\hat{\sigma}^\prime_n}{d\hat{t}}
\times\frac{\langle 0|{\cal O}_n^{H}|0\rangle}{2J+1},
\end{eqnarray}
where
\begin{equation}
\frac{d\hat{\sigma}^\prime_n}{d\hat{t}}
=
\frac{1}{16\pi\hat{s}^2}
\overline{\sum}
|{\cal M}^\prime
\left(
a(p_1)+b(p_2)\rightarrow (Q\overline{Q})_n(P)+c(p_3)
\right)|^2.
\end{equation}
Here, ${\cal M}^\prime$ is the amplitude  of the process
\begin{equation}
a(p_1)+b(P_2)\rightarrow (Q\overline{Q})_n(P)+c(p_3),
\end{equation}
which can be obtained by integrating 
the free particle amplitude over the relative momentum of the 
quark inside the intermediate state $(Q\overline{Q})_n(P)$,
after projecting appropriate spectroscopic state 
Clebsch-Gordon coefficients. The parameter $C_n$ is defined by
\begin{equation}
C_n=\left\{
\begin{array}{cl}
2N_c    &({\rm color-singlet}),\\
N_c^2-1 &({\rm color-octet}).
\label{eq:Cn}
\end{array}
\right.
\end{equation}
And $\langle 0|{\cal O}_n^{H}|0\rangle$
is the non-perturbative matrix element 
representing the transition
\begin{equation}
(Q\overline{Q})(^{2S+1}L_J^{(1,8)}) \to H.
\end{equation}
Finally, $J$ denotes the angular momentum of 
the intermediate state 
$(Q\overline{Q})(^{2S+1}L_J^{(1,8)})$,
not of the physical state $H$.
For the case of color-singlet intermediate state,
which has the same spectroscopic configuration with $H$,
we can relate the matrix elements to the radial wave-function
of the bound state as
\begin{equation}
\frac{\langle 0|{\cal O}_n^{H}|0\rangle}{C_n\times(2J+1)}
=\left\{
\begin{array}{l}
\frac{1}{4\pi}|R_S(0)|^2~~~(S-{\rm wave}),\\
\\
\frac{3}{4\pi}|R_P^\prime(0)|^2~~~(P-{\rm wave}).
\end{array}
\right.
\end{equation}
For example, if we consider the $\psi$ production
via
$(^3S_1^{(1)})$,
$(^1S_0^{(8)})$, $(^3S_1^{(8)})$, $(^3P_0^{(8)})$,
$(^3P_1^{(8)})$ and  $(^3P_2^{(8)})$
intermediate states, then the
partonic subprocess cross sections are given by
\begin{eqnarray}
\frac{d\hat{\sigma}}{d\hat{t}}({\rm octet})
&=&\frac{1}{8M_c}
\left(
\frac{d\hat{\sigma}^\prime}{d\hat{t}}(^1S_0^{(8)})
\times
\langle 0|{\cal O}^{\psi}(^1S_0^{(8)})|0\rangle
+\frac{\hat{d\sigma}^\prime}{d\hat{t}}(^3S_1^{(8)})
\times
\frac{\langle 0|{\cal O}^{\psi}(^3S_1^{(8)})|0\rangle}{3}
\right.
\nonumber\\
&&
\left.
\hskip 2cm
+
\langle 0|{\cal O}^{\psi}(^3P_0^{(8)})|0\rangle
\times
\sum_{J}
\frac{d\hat{\sigma}^\prime}{d\hat{t}}(^3P_J^{(8)})
\right),
\nonumber\\
\frac{d\hat{\sigma}}{d\hat{t}}({\rm singlet})
&=&\frac{1}{M_c}~\frac{|R_S(0)|^2}{4\pi}
 \frac{d\hat{\sigma}^\prime}{d\hat{t}}(^3S_1^{(1)}),
\end{eqnarray}
after imposing the heavy quark spin symmetry
\begin{equation}
\langle 0|{\cal O}^{J/\psi}(^3P_J^{(8)}) |0\rangle
=(2J+1)\langle 0|{\cal O}^{J/\psi}(^3P_0^{(8)}) |0\rangle.
\end{equation}

The subprocess cross section for the color-singlet gluon-gluon 
fusion \cite{berjon} is well known:
\begin{eqnarray}
&&\frac{d\hat{\sigma}}{d\hat{t}}({\rm singlet})
=
\frac{{\cal N}_1}{16\pi \hat{s}^2}~
\left[
\frac{
 \hat{s}^2 (\hat{s}-4M_c^2)^2 
+\hat{t}^2 (\hat{t}-4M_c^2)^2 
+\hat{u}^2 (\hat{u}-4M_c^2)^2} 
{(\hat{s}-4M_c^2)^2 (\hat{t}-4M_c^2)^2 (\hat{u}-4M_c^2)^2 }
\right],
\end{eqnarray}
where the overall normalization ${\cal N}_1$ is defined as
\begin{equation}
{\cal N}_1 = {4 \over 9}~( 4 \pi \alpha_{s})^{2} (4 \pi \alpha) e_{c}^{2}
~M_{c}^{3} G_{1}(J/\psi).
\end{equation}
The parameter $G_1 (J/\psi)$, which is defined in NRQCD as
\begin{equation}
G_{1}( J/\psi) =  
\frac{\langle 0 | {\cal O}(^{3}S_1^{(1)}) | 0  \rangle}{3M_c^2}
=
\frac{3}{2\pi M_c^2}|R_S(0)|^2,
\end{equation}
is proportional to the probability of a color-singlet
$(c \overline{c})$ pair in the $(^{3}S^{(1)}_{1})$ partial wave state 
to form a physical  $J/\psi$ state.  
It is related to the leptonic decay width 
\begin{equation}
\Gamma ( J/\psi \rightarrow l^{+} l^{-} ) = {2 \over 3}~\pi e_{c}^{2}
\alpha^{2}~G_{1}(J/\psi),
\end{equation}
where $e_c = 2/3$. 
{}From the measured leptonic decay rate of $J/\psi$, one can extract
\begin{equation}
G_{1}(J/\psi) \approx 106~~{\rm MeV}.
\end{equation}
After including the radiative corrections of 
${\cal O}(\alpha_{s})$ with $\alpha_{s}(M_{c}) =0.27$, 
this value is increased to $\approx 184$ MeV.
Relativistic corrections tend to increase $G_{1} (J/\psi)$ further to
$\sim 195$ MeV \cite{jungilb}.

For the case of color-octet subprocesses,
we can use the average-squared amplitude of the processes
$\gamma+g~({\rm or}~q)
\rightarrow~(c\overline{c})(^{2S+1}L_J^{(8)})+g~({\rm or}~q)$ 
in Ref. \cite{jungilep}, after crossing
($k\rightarrow -k$~ and~ $q_2\rightarrow -q_2$)
\begin{eqnarray}
\overline{\sum}
|{\cal M}^\prime|^2
(g+g~&&\rightarrow ~(c\overline{c})(^{2S+1}L_J^{(8)})+\gamma)
(\hat{s},\hat{t},\hat{u})\nonumber\\&&=
\frac{1}{8}
|{\cal M}^\prime|^2
(\gamma+g\rightarrow ~(c\overline{c})(^{2S+1}L_J^{(8)})+g)
(\hat{t},\hat{s},\hat{u}),\\
\overline{\sum}
|{\cal M}^\prime|^2
(q+\overline{q}~&&\rightarrow ~(c\overline{c})(^{2S+1}L_J^{(8)})+\gamma)
(\hat{s},\hat{t},\hat{u})\nonumber\\&&=
\frac{1}{3}
|{\cal M}^\prime|^2
(\gamma+q\rightarrow ~(c\overline{c})(^{2S+1}L_J^{(8)})+q)
(\hat{t},\hat{s},\hat{u}).
\end{eqnarray}

As previously explained, there is only one color-singlet subprocess,
but 9 color-octet subprocesses are  
contributing to the process 
\begin{equation}
p + p~(\overline{p}) \rightarrow J/\psi+\gamma+X.
\end{equation}
Whereas the initial partons for the color-singlet process 
are only the gluons,
quarks can also be the initial partons
for the color-octet subprocesses.
For the $(^3S_1^{(8)})$ channel, the
gluon contribution corresponding to Fig. \ref{fig:octet22} (a) is absent
because the effective vertex corresponding to Fig. \ref{fig:agp}
is vanishing.
Using the parton level differential cross sections
for various channels, in the next section
we discuss in detail the
$p_{_T}$ and $M_{J/\psi+\gamma}$ distributions
in the hadronic $J/\psi+\gamma$ production. We also
compare the color-octet contributions with the color-singlet gluon-gluon
fusion contribution.

\section{Numerical Results and Conclusions}

Now, we  are ready to show the numerical results from the analytic
expressions obtained in the previous section.  
The  results are quite sensitive to the numerical values of
QCD coupling constant $\alpha_s$, mass of charm quark $M_{c}$,
and the factorization scale $Q$.
For numerical predictions, we  used
$\alpha_{s} (M_{c}^{2}) = 0.27$, $M_{c} = 1.48$ GeV and
$Q^{2} = (M_T/2)^2$, where $M_T$ is the transverse mass of the outgoing
$J/\psi$. 
For the structure functions, we have used 
the most recent ones, MRSA \cite{mrsa} and CTEQ3M \cite{cteq3},
and it turns out that both sets of structure functions give more or less
the same results within $\sim 10 \%$.

We use the numerical value of the color-octet matrix element,
shown in Eq. (2.12),
\begin{eqnarray}
\langle 0|{\cal O}^\psi(^3S_1^{(8)})|0 \rangle
=(6.6\pm 2.1)\times 10^{-3} {\rm GeV}^3.
\end{eqnarray}
And since the matrix elements 
$ \langle 0|{\cal O}_{8}^{\psi}({^1S_0})|0\rangle$
and 
$ \langle 0|{\cal O}_{8}^{\psi}({^3P_0})|0\rangle$
are not determined separately,
we present the two extreme values allowed by Eqs. (2,13,2.14,\ref{eq:max}) as
\begin{eqnarray}
^1S_0{\rm~~ saturated~~ case}&:&
\langle 0|{\cal O}_{8}^{\psi}({^1S_0})|0\rangle
=(6.6 \pm 1.5)\times 10^{-2}~{\rm GeV}^3,\nonumber\\
&&\langle 0|{\cal O}_{8}^{\psi}({^3P_0})|0\rangle=0,
\nonumber\\
^3P_J{\rm~~ saturated~~ case}&:&
\langle 0|{\cal O}_{8}^{\psi}({^3P_0})|0\rangle
=(2.2 \pm 0.5) \times 10^{-2}~{\rm GeV}^3\times M_{c}^2,\nonumber\\
&&\langle 0|{\cal O}_{8}^{\psi}({^1S_0})|0\rangle=0.
\label{eq:over}
\end{eqnarray}
After analyzing the numerical results, we found that the color-octet
contribution of $(^3S_1^{(8)})$  was significantly suppressed 
compared to the others over the entire phase space. 
Therefore, we do not present the result for the channel $(^3S_1^{(8)})$.


We first consider the
$M_{J/\psi+\gamma}$ distribution at a fixed target experiment.
In Ref. \cite{psigamma_mass}, Berger and Sridhar 
investigated the $J/\psi+\gamma$ invariant mass distribution at 
ISR \cite{isr} and E705 \cite{e705} experiments. They argue that 
this process forms an important (and computable) part of the
background to the decay of $P$-state charmonia into $J/\psi+\gamma$ 
at low values of the invariant mass of the $J/\psi + \gamma$ pair.
In Fig. \ref{fig:mass}, we present the $M_{J/\psi+\gamma}$ distribution 
for the E705  fixed target experiment ($\sqrt{s}=23.72$ GeV).
We note that the value of  $M_{J/\psi+\gamma}$, to produce the 
$J/\psi+\gamma$ pair, must be larger than $M_{J/\psi}$.
The solid line is the color-singlet gluon-gluon fusion contribution.
The dashed line represents $^3P_J^{(8)}$-saturated color-octet contribution,
and the dotted one is  $^1S_0^{(8)}$-saturated color-octet contribution.
As shown in this figure, the color-octet  contributions are strongly
suppressed compared to that of the color-singlet gluon-gluon fusion process.
Considering the fact that the 
$^3P_J^{(8)}$-saturated and $^1S_0^{(8)}$-saturated results
are maximally allowed (upper bound) contributions, 
as explained in Eqs. (2.14,2.15,3.2),
the color-octet contributions must be smaller than the color-singlet one
by more than one order of magnitude all over the region.


In Ref. \cite{frag_rs} the color-singlet gluon-gluon fusion contribution 
has been compared with the gluon and charm-quark fragmentations 
to inclusive $J/\psi$ production at Tevatron energies. It was found that 
the fragmentation contributions are more than an order of magnitude 
smaller than the color-singlet gluon
fusion contribution over the whole range of $p_{_T}$ considered.
In Fig. \ref{fig:pt}, we show $B d\sigma /d p_{_T}$ for
$J/\psi+\gamma$ production in $p \overline p$ collisions 
at $\sqrt{s}=1.8$~TeV, integrated over a rapidity interval 
$\vert y_{_{J/\psi}} \vert \le 0.5$, where $B$ is the $J/\psi$
branching ratio into leptons ($B$=0.0594).
All the line styles in Fig. 7 are identical with those of 
Fig. \ref{fig:mass}.
As shown in Fig.~\ref{fig:pt}, the
color-singlet component is dominant where $p_{_T}<6$~GeV.
There is a cross over point where color-octet components dominate
that of color-singlet gluon-gluon fusion subprocess
around $p_{_T}=6$~GeV.
Though the color-octet contributions dominate in the high $p_{_T}$ region,
the cross section in this region is much smaller than that of low $p_{_T}$ 
region, where the color-singlet contribution dominates. 
As previously explained, the 
$^3P_J^{(8)}$-saturated and $^1S_0^{(8)}$-saturated results
are maximally allowed (upper bound) contributions, and
the color-octet contributions might be smaller than the color-singlet one
even at large $p_{_T}>6$ GeV.


In Fig.~\ref{fig:ym} 
we show the rapidity distribution of $J/\psi$, 
$B d \sigma /dy$, integrated over the 
transverse momentum of $J/\psi$, $3~{\rm GeV}<p_{_T}<{\rm 6~GeV}$.
The rapidity distribution integrated over this $p_{_T}$ region
is dominated by the color singlet gluon fusion process, as shown in 
Fig.~\ref{fig:pt}. Considering the fact that the matrix elements
$\langle0|{\cal O}^{J/\psi}(^1S_0)|0\rangle$ and 
$\langle0|{\cal O}^{J/\psi}(^3P_0)|0\rangle$ in Eq.~({\ref{eq:over}) 
are possibly overestimated by about an order of magnitute,
the color octet contributions shown in Fig.~\ref{fig:ym}
could be negligible compared to the color singlet gluon
fusion process. Finally, we note that
even though we have used the covariant projection method 
in our calculation, we have considered only the unpolarized $J/\psi$
production.
Therefore, there are no interference terms among different
$^3P_J^{(8)}$ components, and our results shown here are perfectly
valid~\cite{bra_chen,bene_roth}.


To conclude, we have considered the color-octet contributions to the associate 
$J/\psi+\gamma$ production in the hadronic collisions, also
compared to the conventional color-singlet gluon-gluon fusion contribution.
Within the color-singlet model the $J/\psi+\gamma$ production in the leading 
order is possible only through gluon-gluon fusion process.
As we expected according to the naive power counting, we found that 
the color-octet contributions are significantly suppressed compared
to the color-singlet gluon-gluon fusion process.
For the case of the hadroproduction at the Tevatron energies,
we found that there is a cross-over point at high $p_{_T}$ 
where the color-octet contributions become dominant,
and it is still larger than the fragmentation contributions
shown in Ref. \cite{psigamma_frag}.
This suppression of the color-octet contributions in the associated 
production is qualitatively consistent with the recent result of Cacciari 
and Kr\"{a}mer~\cite{kramer_new}. They investigated the  
$J/\psi+\gamma$ production via resolved photon collisions in HERA, and
found the strong suppression on the color-octet contributions compared to
the color-singlet one as
at $\sqrt{s_{\gamma p}}=100$~GeV for $p_T>1$~GeV~\cite{kramer_new},
\begin{eqnarray}
\frac{\sigma(^1S_0^{(8)})+\sigma(^3P_J^{(8)})}
{\sigma(^3S_1^{(1)})}&\approx&15\%~.
\end{eqnarray}
Though most events at the Tevatron collider are in the region 
where the color-singlet gluon-gluon fusion contribution dominates, 
one could require an additional cut, for example:  $p_{_T}<6$~GeV,
to guarantee that the color-singlet gluon-gluon fusion process remains 
one of the cleanest channels to probe the gluon distribution
inside proton, and to study  heavy quarkonia production mechanism.


\acknowledgements
The authors would like to express special gratitude to Ed Berger, M. Drees
and Pyungwon Ko for critical discussions and comments on the manuscript. 
The authors also would like to thank M. Cacciari, M. Greco and M. Kr\"amer
for pointing out the misprints in the labelling of the figures in the
original version.\footnote{ After submitting this paper, there appeared 
the paper\cite{Cacci} showing the consistent result with ours.}
The work was supported in part by the Korea Science and Engineering
Foundation through the SRC program and in part by Korea Research Foundation.
The work of CSK was supported 
in part by the Korea Science and Engineering  Foundation, 
Project No. 951-0207-008-2,
in part by the Basic Science Research Institute Program,
Ministry of Education 1997,  Project No. BSRI-97-2425, and
in part by the Code of Excellence Fellowship from Japanese Ministry of 
Education, Science and Culture.


\newpage
\begin{center}
FIGURE CAPTIONS
\end{center}
\noindent
Fig.1
\hskip .3cm
{Feynman diagrams for the color-singlet subprocess
for $g + g \rightarrow (c\overline{c})(^{3}S_{1}^{(1)}) + \gamma$
and the color-octet subprocess
for $g + g \rightarrow 
(c\overline{c})(^{1}S_{0}^{(8)}~{\rm or}~^{3}P_{J}^{(8)}) + \gamma$.
}
\\
\\
Fig.2
\hskip .3cm
{Feynman diagrams for the effective 
$\gamma-g-(c\overline{c})(^1S_0^{(8)},~ {\rm or }~^3P_J^{(8)})$ vertex.}
\\
\\
Fig.3
\hskip .3cm
{Feynman diagram for the effective 
$q-\overline{q}-(c\overline{c})(^{3}S_{1}^{(8)})$ vertex.}
\\
\\
Fig.4
\hskip .3cm
{Feynman diagrams for the color-octet contributions to the subprocesses
(a) $g+g \rightarrow 
(c\overline{c})(^1S_0^{(8)},~{\rm or }~^3P_J^{(8)})+\gamma$~ 
and  (b) $q+ \overline{q} \rightarrow
(c\overline{c})(^1S_0^{(8)},~ {\rm or }~^3P_J^{(8)})+\gamma$.}
\\
\\
Fig.5
\hskip .3cm
{Feynman diagrams for the color-octet contributions to the
subprocess $ q  + \overline{q} \rightarrow
(c\overline{c})(^3S_1^{(8)}) +\gamma$.}
\\
\\
Fig.6
\hskip .3cm
{Invariant mass $M_{J/\psi+\gamma}$ distribution
of the process $p p \to J/\psi+\gamma$, where $\sqrt{s}=23.72$GeV.
The solid line is the color-singlet gluon-gluon fusion contribution.
The dashed line represents the 
$^3P_J^{(8)}$-saturated color-octet contribution,
and the dotted one is the $^1S_0^{(8)}$-saturated color-octet contribution.
}
\\
\\
Fig.7
\hskip .3cm
{Transverse momentum of $J/\psi$, $p_{_T}(J/\psi)$, distribution
$B d\sigma/dp_{_T}$, integrated over the $J/\psi$ rapidity range
$|y_{_{J/\psi}}|<0.5$, for the process $p\overline{p}\to J/\psi+\gamma+X$,
where $\sqrt{s}=1.8$~TeV.
Here, $B$ is the branching ratio of $J/\psi$ decay into leptons ($B$=0.0594).
The solid line represents the color-singlet gluon-gluon fusion contribution.
The dashed line represents the 
$^3P_J^{(8)}$-saturated color-octet contribution,
and the dotted one is the $^1S_0^{(8)}$-saturated color-octet contribution.
}
\\
\\
Fig.8
\hskip .3cm
{Rapidity of $J/\psi$, $y(J/\psi)$, distribution $B d\sigma/dy_{_{J/\psi}}$, 
integrated over the $J/\psi$ transverse momentum range
$3~{\rm GeV}<p_{_T}<{\rm 6~GeV}$, 
for the process $p\overline{p}\to J/\psi+\gamma+X$,
where $\sqrt{s}=1.8$~TeV. 
Here, $B$ is the branching ratio of $J/\psi$ decay into leptons ($B$=0.0594).
The solid line represents the color-singlet gluon-gluon fusion contribution.
The dashed line represents the 
$^3P_J^{(8)}$-saturated color-octet contribution,
and the dotted one is the $^1S_0^{(8)}$-saturated color-octet contribution.
}
\begin{figure}
\vskip 2cm
\hbox to\textwidth{\hss\epsfig{file=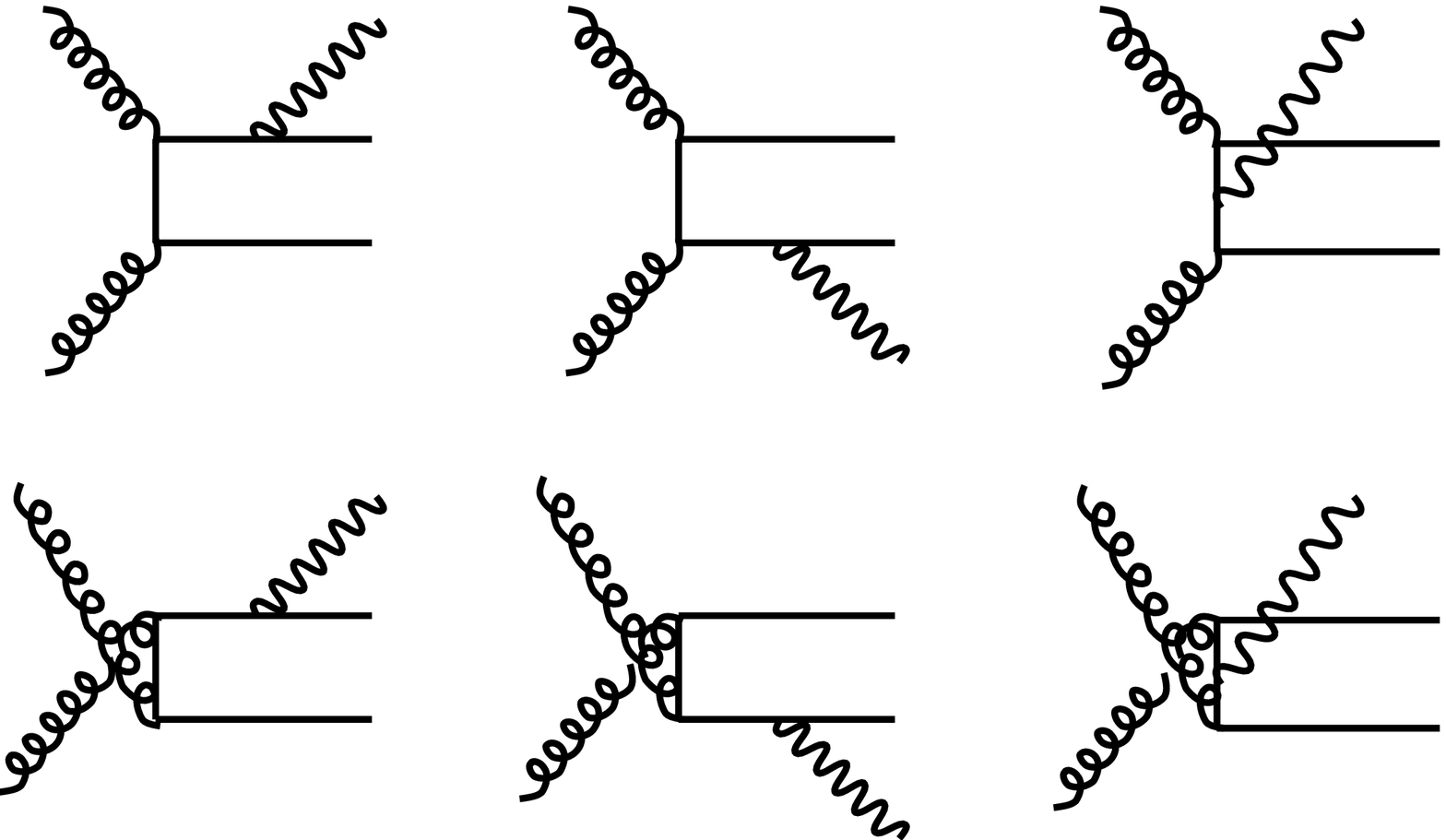,width=13cm}\hss}
\vskip 0.5cm
\caption{}
\label{fig:fusion}
\end{figure}
\begin{figure}
\vskip 2cm
\hbox to\textwidth{\hss\epsfig{file=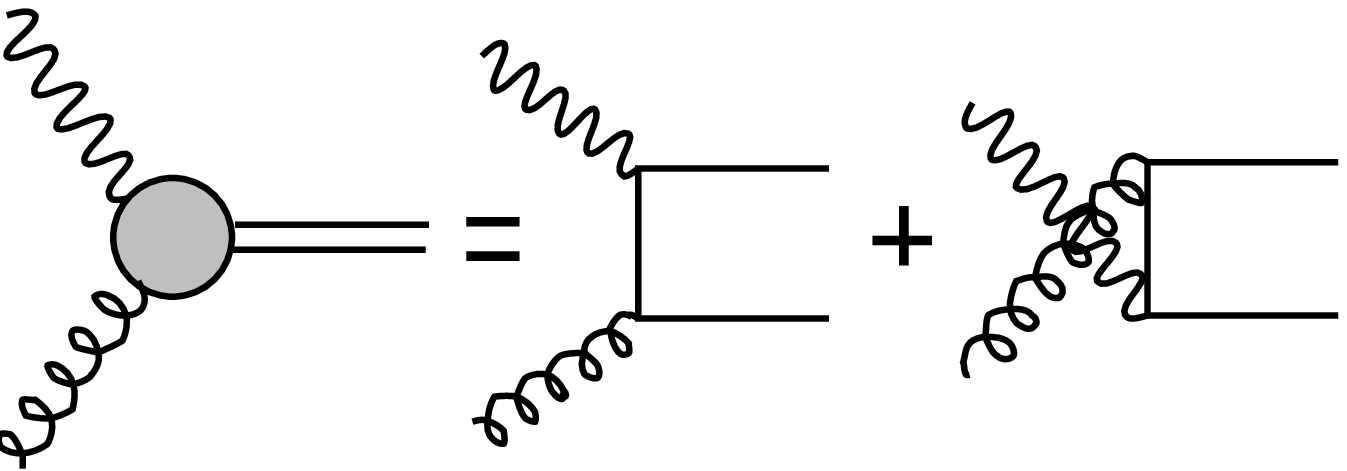,width=13cm}\hss}
\vskip 0.5cm
\caption{}
\label{fig:agp}
\end{figure}
\newpage
\begin{figure}
\vskip 4cm
\hbox to\textwidth{\hss\epsfig{file=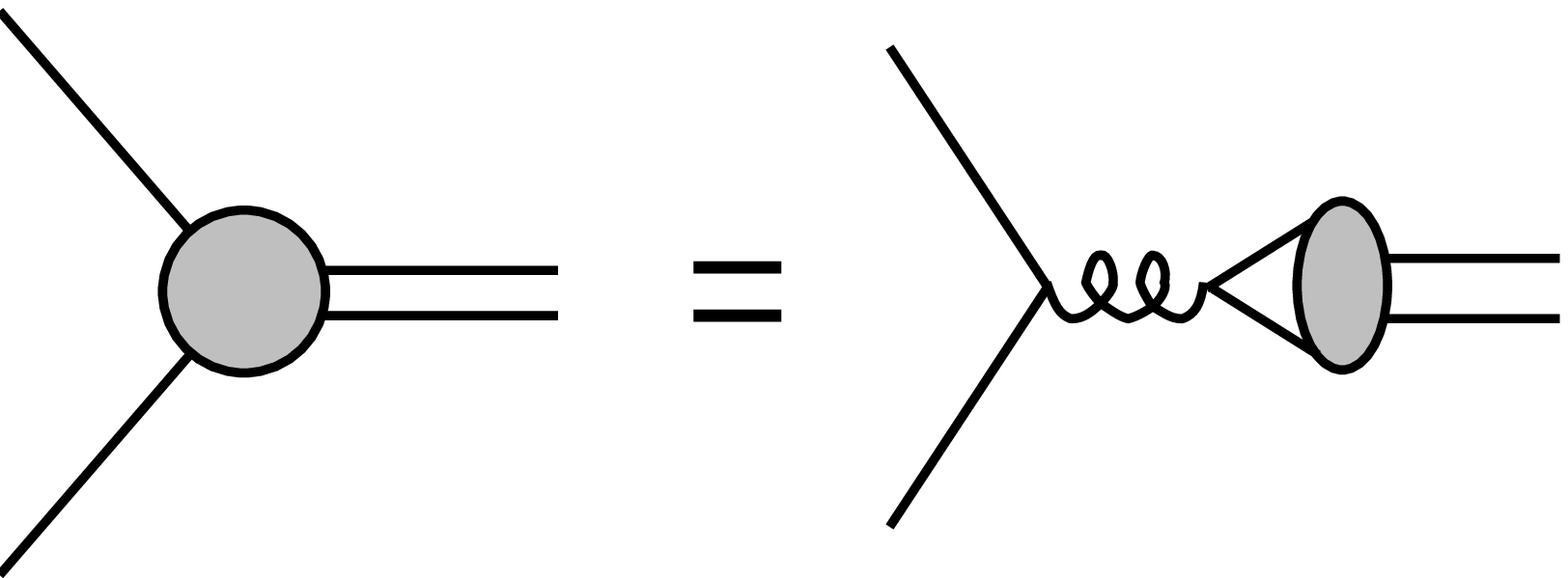,width=13cm}\hss}
\vskip 0.5cm
\caption{}
\label{fig:qqp}
\end{figure}
\begin{figure}
\vskip 2cm
\hbox to\textwidth{\hss\epsfig{file=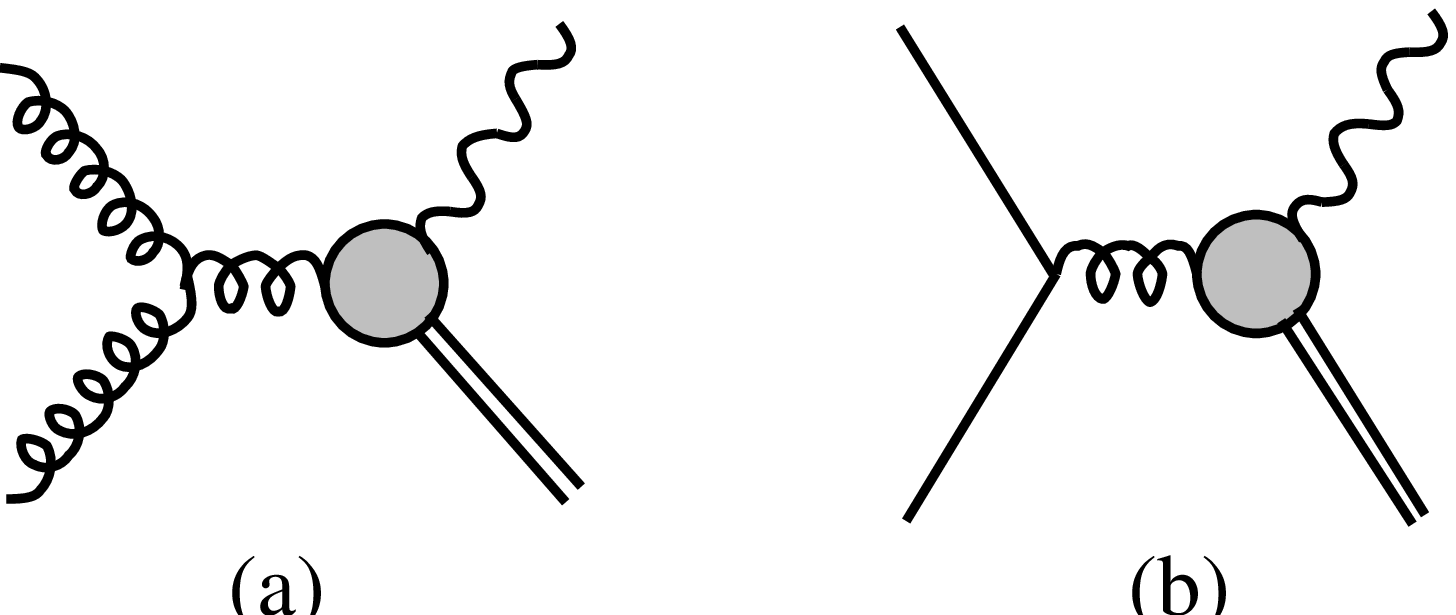,width=13cm}\hss}
\vskip 1.cm
\caption{}
\label{fig:octet22}
\end{figure}
\newpage
\begin{figure}
\vskip 5cm
\hbox to\textwidth{\hss\epsfig{file=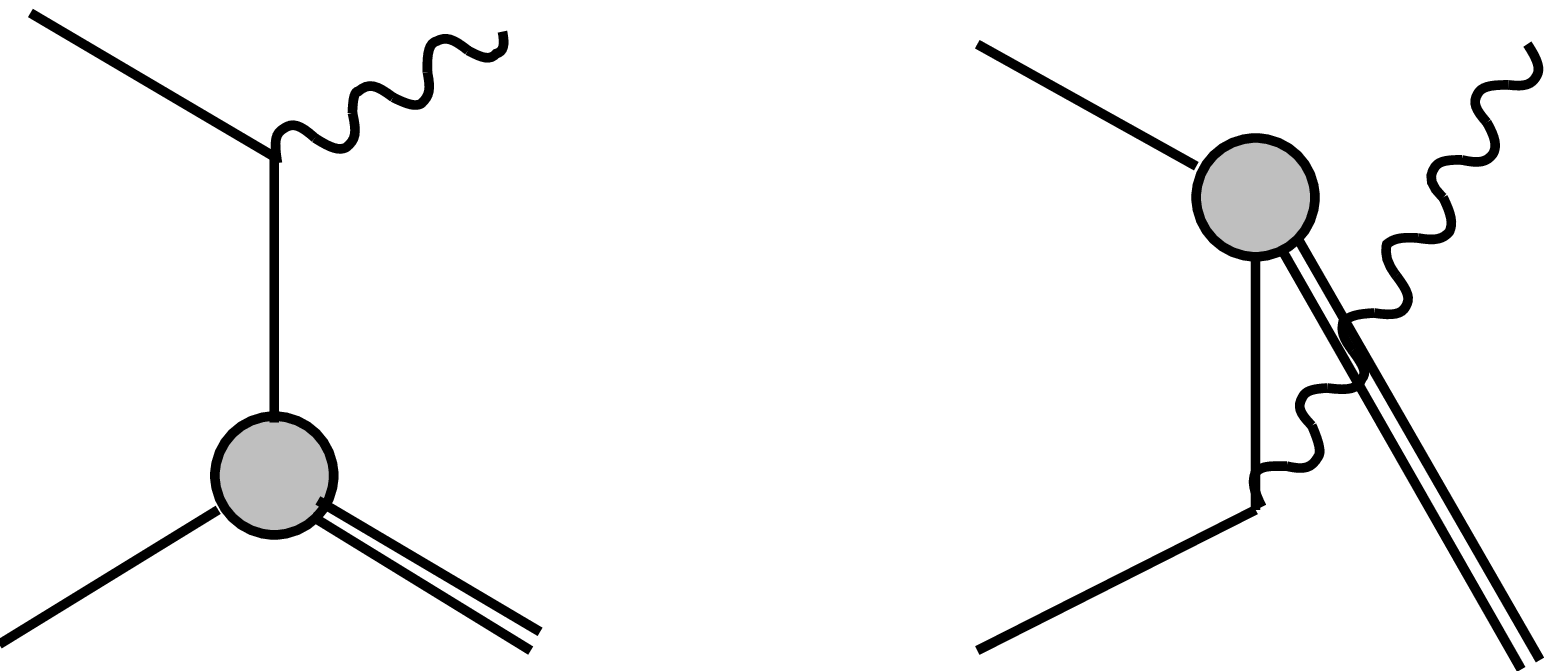,width=13cm}\hss}
\vskip 1.cm
\caption{}
\label{fig:qq3s1}
\end{figure}
\newpage
\begin{figure}
\vskip 5cm
\hskip -1.5cm
\hbox to\textwidth{\hss\epsfig{file=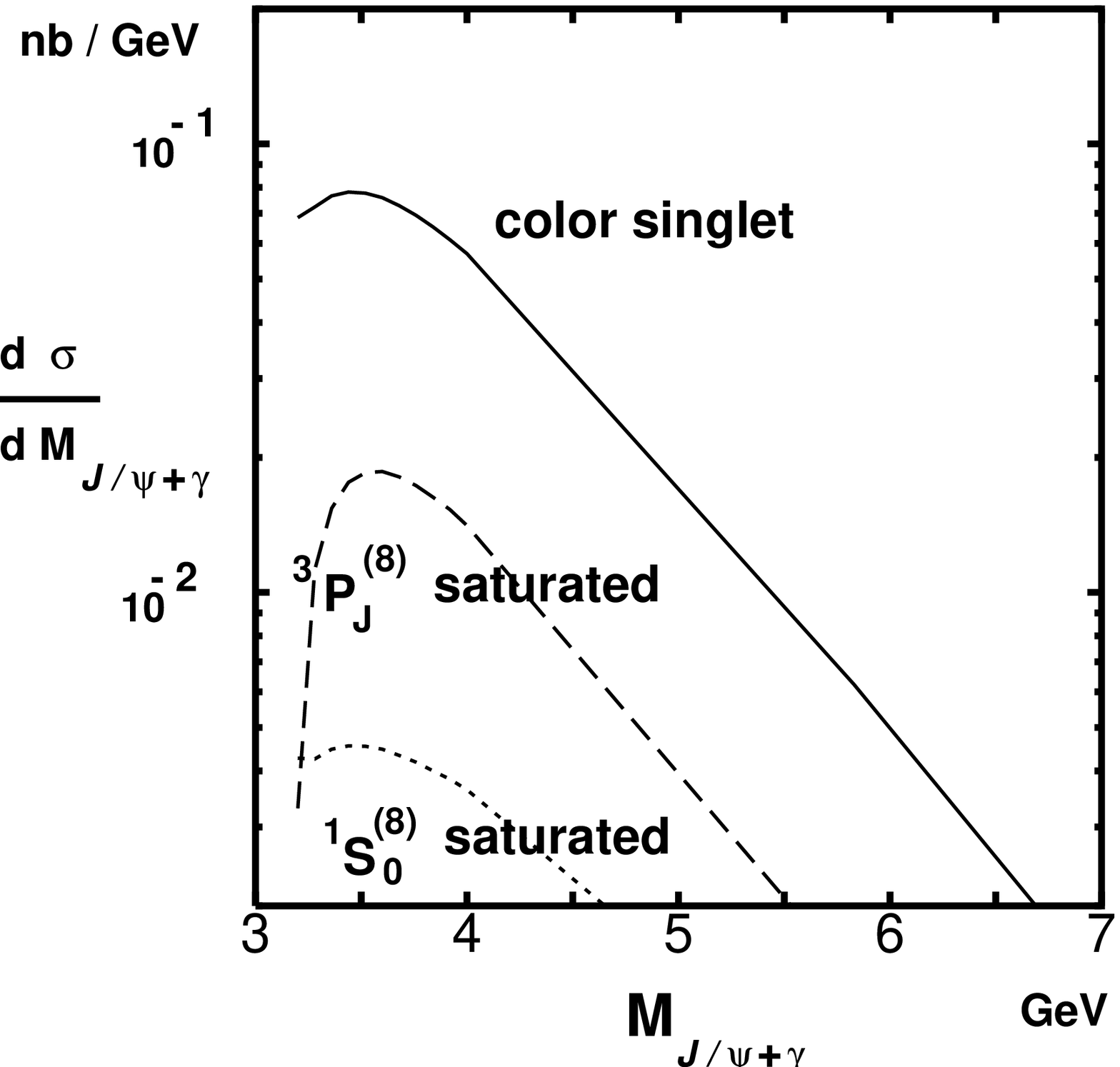,width=15cm}\hss}
\vskip 1.cm
\caption{}
\label{fig:mass}
\end{figure}
\begin{figure}
\vskip 5cm
\hskip -1.5cm
\hbox to\textwidth{\hss\epsfig{file=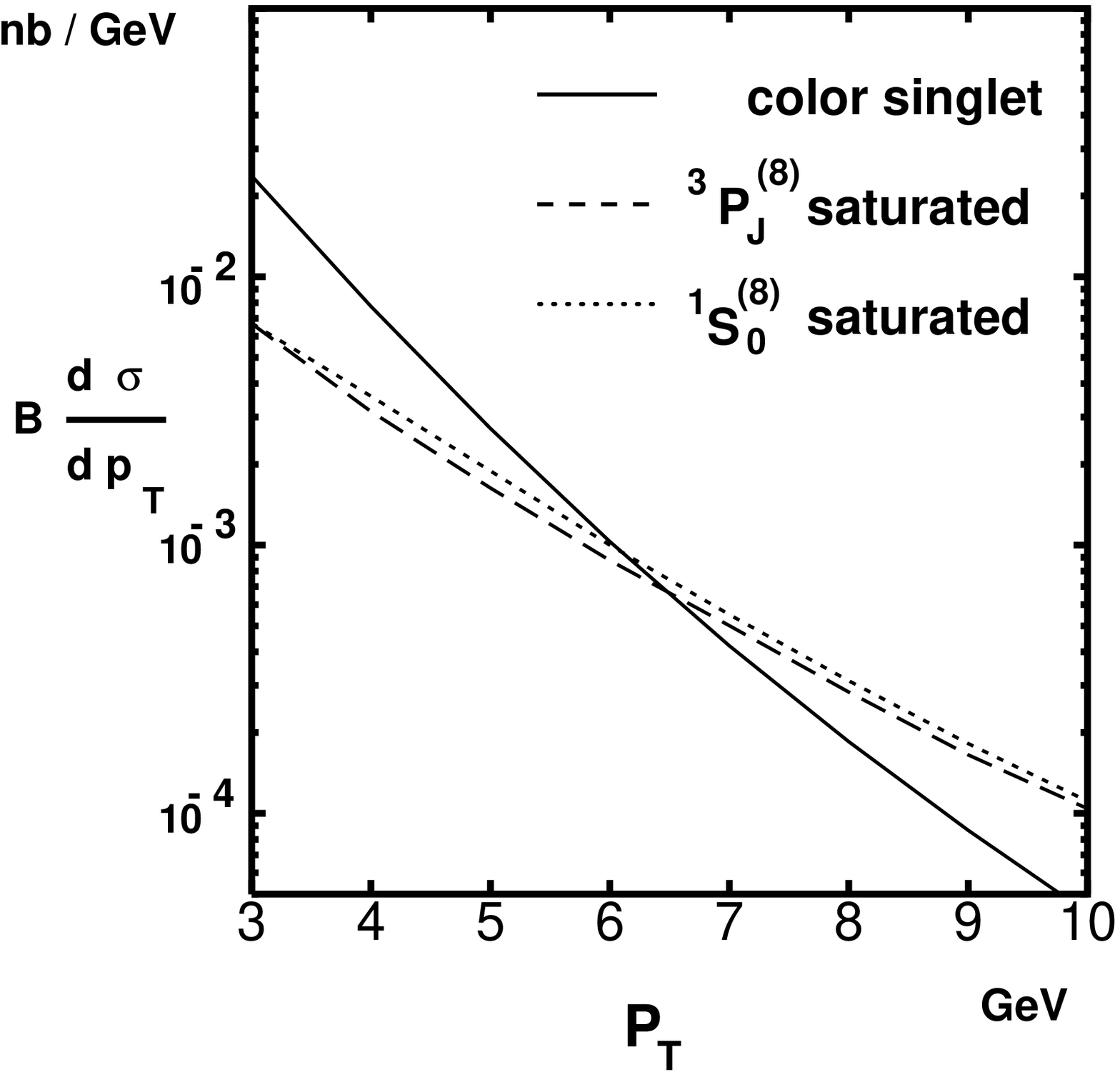,width=15cm}\hss}
\vskip 1.cm
\caption{}
\label{fig:pt}
\end{figure}
\begin{figure}
\vskip 5cm
\hskip -1.5cm
\hbox to\textwidth{\hss\epsfig{file=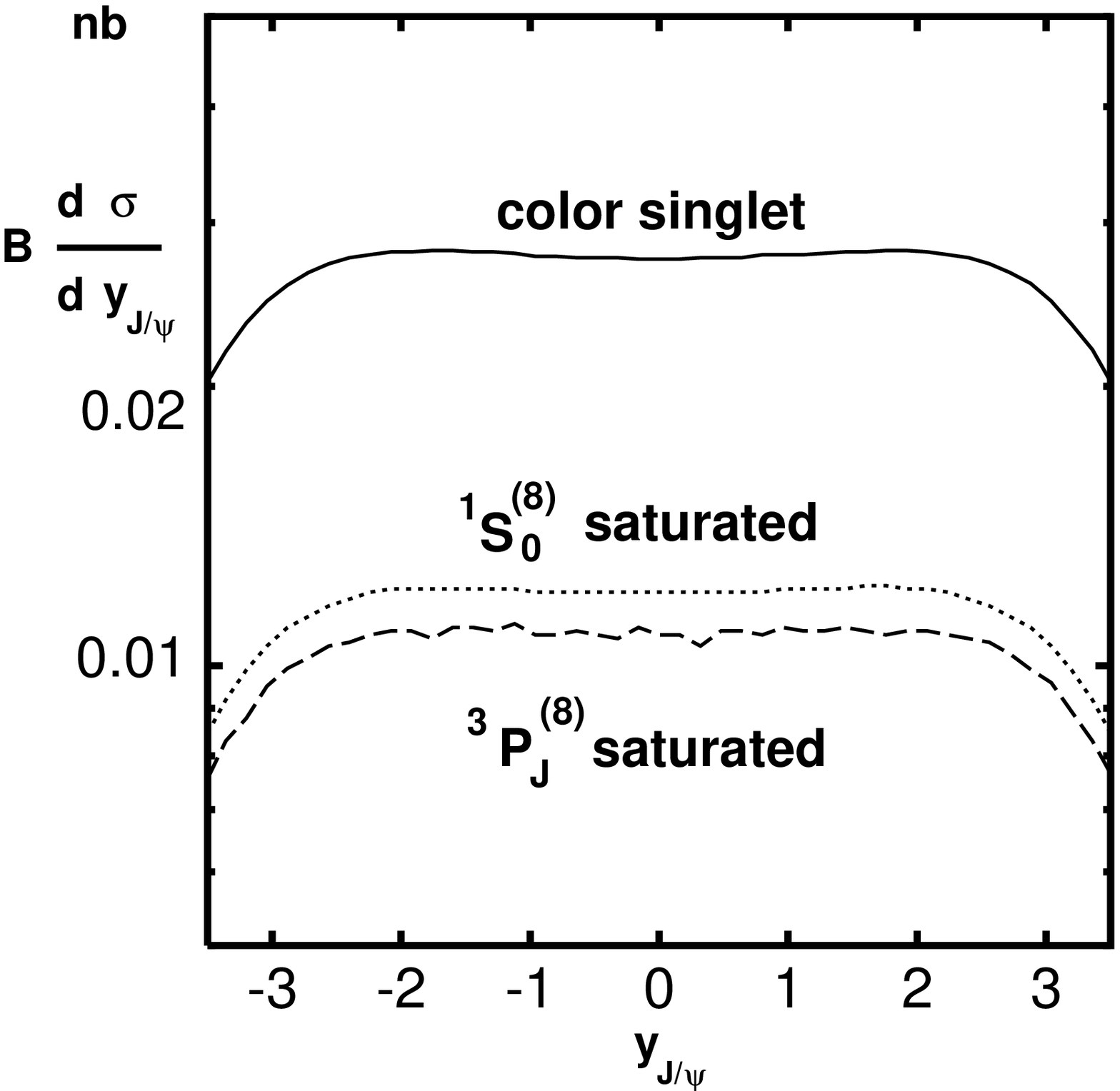,width=15cm}\hss}
\vskip 1.cm
\caption{}
\label{fig:ym}
\end{figure}
\end{document}